\pdfoutput = 1
\documentclass[10pt, letterpaper, conference, twocolumn]{IEEEtran}

\usepackage{amsmath}
\usepackage{amssymb}
\usepackage{siunitx}
\DeclareSIUnit\dBm{dBm}
\usepackage{bm}
\usepackage{adjustbox}
\usepackage[caption=false,font=footnotesize]{subfig}
\usepackage{commath}
\newtheorem{theorem}{Theorem}
\newtheorem{remark}{Remark}
\newtheorem{proposition}{Proposition}

\newcommand{\E}[2]{\mathbb{E}_{#2}\!\left[#1\right]}
\newcommand{\vect}[1]{\bm{#1}}

\newcommand{\trans}{^{\mathrm{T}}}
\newcommand{\herm}{^{\mathrm{H}}}
\newcommand{\e}{\mathrm{e}}

\DeclareMathOperator{\jj}{j}

\DeclareMathOperator*{\trace}{tr}

\DeclareMathOperator*{\argmax}{argmax}

\DeclareMathOperator*{\atan2}{atan2}
\DeclareMathOperator*{\supp}{supp}

\usepackage{tikz}
\usepackage{pgfplots}
\pgfplotsset{compat=newest}
\usepgfplotslibrary{colormaps}
\usetikzlibrary{arrows.meta, backgrounds, calc, decorations, decorations.markings, decorations.pathreplacing, math, matrix, patterns, pgfplots.polar, shapes}
\usepackage{pgfplotstable}

\title{5G Downlink Multi-Beam Signal Design for LOS Positioning}
\author{
	\IEEEauthorblockN{Anastasios~Kakkavas\IEEEauthorrefmark{1}\IEEEauthorrefmark{2}, Gonzalo~Seco-Granados\IEEEauthorrefmark{3}, Henk~Wymeersch\IEEEauthorrefmark{4}, Mario~H.~Casta\~neda~Garc\'ia\IEEEauthorrefmark{1}, \\Richard A. Stirling-Gallacher\IEEEauthorrefmark{1} and Josef~A.~Nossek\IEEEauthorrefmark{2}\IEEEauthorrefmark{5}}
	\IEEEauthorblockA{\IEEEauthorrefmark{1}Munich~Research~Center,~Huawei~Technologies~Duesseldorf~GmbH,~Munich,~Germany}
	\IEEEauthorblockA{\IEEEauthorrefmark{2}Department~of~Electrical~and~Computer Engineering, Technische Universit{\"{a}}t M{\"{u}}nchen, Munich, Germany}
	\IEEEauthorblockA{\IEEEauthorrefmark{3}Department~of~Telecommunications~and~Systems~Engineering,~Universitat Autonoma~de~Barcelona,~Spain}
	\IEEEauthorblockA{\IEEEauthorrefmark{4}Department~of~Electrical~Engineering,~Chalmers~University~of~Technology,~Sweden}
	\IEEEauthorblockA{\IEEEauthorrefmark{5}Department~of~Teleinformatics~Engineering,~Federal~University~of~Cear{\'a},~Fortaleza,~Brazil}
}

\usepackage{textcomp}
\newcommand\copyrighttext{%
	\footnotesize \textcopyright 2019 IEEE. Personal use of this material is permitted. Permission from IEEE must be obtained for all other uses, in any current or future media, including reprinting/republishing this material for advertising or promotional purposes, creating new collective works, for resale or redistribution to servers or lists, or reuse of any copyrighted component of this work in other works.}

\newcommand\copyrightnotice{%
	\begin{tikzpicture}[remember picture,overlay]
	\node[anchor=south,yshift=10pt] at (current page.south) {{\parbox{\dimexpr\textwidth-\fboxsep-\fboxrule\relax}{\copyrighttext}}};
	\end{tikzpicture}%
}

\newcommand\conferenceinfotext{%
	\footnotesize A. Kakkavas, G. Seco-Granados, H. Wymeersch, M. H. Casta\~neda Garc\'ia, R. A. Stirling-Gallacher and J. A. Nossek, "5G Downlink Multi-Beam Signal Design for LOS Positioning," IEEE Global Communications Conference (GLOBECOM), Waikoloa, HI, USA, 2019, pp. 1-6.}

\newcommand\conferenceinfonotice{%
	\begin{tikzpicture}[remember picture,overlay]
		\node[anchor=south,yshift=-35pt] at (current page.north) {{\parbox{\dimexpr\textwidth-\fboxsep-\fboxrule\relax}{\conferenceinfotext}}};
	\end{tikzpicture}%
}

\usepackage[pdfencoding=auto, psdextra]{hyperref}
\usepackage{bookmark}
\hypersetup{
	pdfinfo={
		Title={5G Downlink Multi-Beam Signal Design for LOS Positioning},
		Author={Anastasios Kakkavas, Gonzalo Seco-Granados, Henk Wymeersch,  Mario H. Casta\~neda Garc\'ia, Richard A. Strirling-Gallacher and Josef A. Nossek}
	}
}

\begin{document}
	\maketitle
	\IEEEoverridecommandlockouts
	\copyrightnotice
	\conferenceinfonotice
	
	\vspace*{-0.35cm}
	\begin{abstract}
		In this work, we study optimal transmit strategies for minimizing the positioning error bound in a line-of-sight scenario, under different levels of prior knowledge of the channel parameters. For the case of perfect prior knowledge, we prove that two beams are optimal, and determine their beam directions and optimal power allocation. For the imperfect prior knowledge case, we compute the optimal power allocation among the beams of a codebook for two different robustness-related objectives, namely average or maximum squared position error bound minimization. Our numerical results show that our low-complexity approach can outperform existing methods that entail higher signaling and computational overhead.
	\end{abstract}

	\section{Introduction}
		Accurate position information is essential in realizing a variety of new use cases of the fifth generation of mobile communication networks (5G), like smart factories~\cite{WHK+16} and automated/assisted driving~\cite{WSD+17}. Many positioning techniques have been developed considering the presence of multiple reference stations (anchors)~\cite{HSZ+16,MB19}. However, multi-anchor positioning might be impossible at millimeter-wave (mm-Wave) frequencies with two or more LoS links to different base stations will be challenging.
		On the other hand, more antennas can be packed in the same area at mm-Wave frequencies~\cite{SAH+14}, enabling high angular resolution. Furthermore, large bandwidths are also available at mm-Wave frequencies, which allow for a high temporal resolution. Based on these, it has been shown that reliable position estimation is indeed possible with a single anchor in line-of-sight (LOS) and non-LOS scenarios~\cite{SGD+15,AZA+18}. 
		
		Single-anchor positioning can be performed in the downlink or uplink \cite{AZA+18}. A common procedure for single-anchor positioning involves sweeping at the transmitter (Tx) a set of beams in a codebook, assuming no prior knowledge of the user's position~\cite{GSK+18,FCW+18}. In several cases, however, prior knowledge of the user's position or channel parameters may be available through the Global Navigation Satellite System, prior training phases, tracking, or known user distributions~\cite{GWS+16}.
		Based on such prior knowledge, optimal beamforming design for single-anchor positioning has been recently treated in the literature. In \cite{SGD+15}, the necessary condition on the reference signal for a non-singular Fisher information matrix (FIM) is derived, concluding that multiple nearby beams provide good conditions for joint estimation of position and orientation in LOS mm-Wave systems. For a LOS multiple-input multiple-output (MIMO) multicarrier system, the optimal transmit beamformer per subcarrier is derived in~\cite{KDD+17}, with known receiver (Rx) position, in order to minimize the delay and angle of arrival (AOA) estimation error. In~\cite{GWS18}, optimal transmit beamformers are derived, so that the worst-case CRLB of the angle of departure (AOD) and AOA of a single path is minimized, under a given uncertainty range for the AOD and AOA. 
		Assuming perfect knowledge of a user's distance and AOD, a transmit beamforming design approach is proposed in \cite{KDU+18}, based on minimizing a weighted sum of the CRLBs of the delay, AOD and AOA estimation of the user's LOS path. Based on this, beamforming optimization in a multi-user setup is considered as well.
		The optimal Rx beamformers for lossless signal compression of a single path are derived in~\cite{ZZS18}, showing that two beams contain all the localization information regardless of the number of antennas.
		
		In this work, we study the optimal beamforming directions for minimizing the positioning error bound (PEB). We prove that, 
		with perfect prior knowledge at the Tx of the AOD and the distance between Tx and Rx, two transmit beams are optimal. For these two beams we derive the optimal beam directions and power allocation.
		Furthermore, we study the case of a fixed beam codebook and compute the optimal power allocation over its beams under perfect and imperfect knowledge of the channel parameters, in order to provide robust positioning. Although we only consider a single user, the discussed approach is naturally applicable to the multi-user case.
		
		The rest of the paper is organized as follows. The system model is derived in Section II and the Cram\'{e}r-Rao bound for position and orientation estimation is derived in Section III. In Section IV the signal design approaches are presented. Numerical evaluations of the proposed methods are provided in Section V and Section VI concludes the work. 
		
	\section{System Model}
		\subsection{Geometry}
\begin{figure}
				\centering
				\begin{adjustbox}{scale = 0.92}
				\includegraphics[]{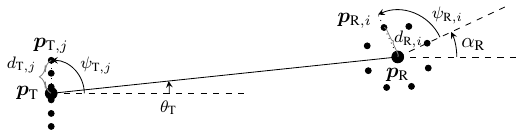}
				\end{adjustbox}
				\caption{Geometric model; example with a uniform linear array (ULA) at the transmitter and a uniform circular array (UCA) at the receiver.}
				\label{fig:geometric_model}
			\end{figure}

			We consider a LOS single-anchor setup, as shown in Fig.~\ref{fig:geometric_model}.
			The Tx and Rx are equipped with antenna arrays consisting of $N_{\text{T}}$ and $N_{\text{R}}$ elements. The arrays' reference points, which are at their centroids, are located at $\vect{p}_{\text{T}} = [0,\; 0]\trans$ and $\vect{p}_{\text{R}} = [p_{\text{R,x}},\; p_{\text{R,y}}]\trans\in\mathbb{R}^2$, where $(\cdot)\trans$ denotes transposition. The Rx's orientation with respect to (w.r.t) the Tx frame of reference is $\alpha_{\text{R}}$. The position of the $j$-th element of the Tx array is $\vect{p}_{\text{T},j} = \vect{p}_{\text{T}} + d_{\text{T},j}\vect{u}\left(\psi_{\text{T},j}\right)$, where $d_{\text{T},j}$ is its distance from $\vect{p}_{\text{T}}$ and $\psi_{\text{T},j}$ is its angle w.r.t. the Tx frame of reference, with $\vect{u}\left(\psi\right) = [\cos\psi,\;\sin\psi]\trans$. The position of the Rx array's elements is defined accordingly. 
			The distance between the Tx and Rx is $d = \left\|\vect{p}_{\text{R}} - \vect{p}_{\text{T}}\right\|_2$, $\theta_{\text{T}} = \atan2\left(p_{\text{R,y}},\; p_{\text{R,x}}\right)$ is the AOD, with $\atan2(\cdot,\cdot)$ being the four-quadrant inverse tangent function, 
			and $\theta_{\text{R}} = \theta_{\text{T}} + \pi - \alpha_{\text{R}}$ is the AOA.
		
		\subsection{Signal Model}
			An Orthogonal Frequency Division Multiplexing (OFDM) waveform, with $N$ subcarriers and a subset $\mathcal{P}$ of them occupied, is used. We assume a narrowband signal model, i.e. $f_{\text{s}} \ll f_{\text{c}}$, where $f_{\text{s}}/N$ is the subcarrier spacing and $f_{\text{c}}$ is the carrier frequency. In order to simplify the presentation, we consider a single OFDM symbol transmission, but the analysis can be straightforwardly extended to the multi-symbol case. The Tx-Rx clock synchronization error $\epsilon_{\text{clk}}$ is modeled as a zero-mean Gaussian random variable with variance $\sigma_{\text{clk}}^2$. The received signal $\vect{y}[p]\in\mathbb{C}^{N_{\text{R}}}$ at the $p$-th subcarrier is
			\begin{IEEEeqnarray}{rCl}
				\vect{y}[p] &=& \vect{m}[p] + \vect{\eta}[p],\\
				\vect{m}[p] &=& h \e^{-\jj \omega_p \tau} \vect{a}_{\text{R}}\left(\theta_{\text{R}}\right) \vect{a}_{\text{T}}\trans\left(\theta_{\text{T}}\right) \vect{x}[p],
				\label{eq:signal model}
				\IEEEeqnarraynumspace
			\end{IEEEeqnarray}
			where $h$ is the channel gain, $\tau = d/c + \epsilon_{\text{clk}}$ is the observed time of flight, with $c$ being the speed of light, $\vect{x}[p]\in\mathbb{C}^{N_{\text{T}}}$ is the Tx signal at the $p$-th subcarrier, $\vect{\eta}[p]$ is the spatially and temporally white Gaussian noise with variance $\sigma_{\eta}^2$ and $\omega_p = 2\pi f_{\text{s}} p/N$. The $j$-th element of the Tx array steering vector $\vect{a}_{\text{T}}\left(\theta_{\text{T}}\right)\in\mathbb{C}^{N_{\text{T}}}$ is defined as $a_{\text{T},j}\left(\theta_{\text{T}}\right) = \e^{\jj \omega_c d_{\text{T},j}\vect{u}\trans\left(\psi_{\text{T},j}\right)\vect{u}\left(\theta_{\text{T}}\right)/c},\; j=1,\ldots,N_{\text{T}}$, with $\omega_{\text{c}} = 2\pi f_{\text{c}}$. The Rx steering vector $\vect{a}_{\text{R}}\left(\theta_{\text{R}}\right)\in\mathbb{C}^{N_{\text{R}}}$ is defined similarly. In addition, the Tx uses an exponential path loss model for the large scale fading of the channel gain: $|h|^2 = |h_0|^2(d_0/d)^n$, where $n\geq 0$ is the path loss exponent, $h_0$ is the reference value of the gain at distance $d_0$. Due to the fact that the synchronization error has no effect on the following analysis, we assume $\sigma_{\text{clk}} = 0$, but in the numerical results we consider practical values for $\sigma_{\text{clk}}$.
	
	\section{CRLB for Position and Orientation Estimation}
		Knowning the Tx position $\vect{p}_{\text{T}}$ and using the observations $\vect{y}[p],\;p\in\mathcal{P}$, the receiver aims to estimate its position $\vect{p}_{\text{R}}$ and potentially its orientation $\alpha_{\text{R}}$, which influence the observations through the channel parameters $\tau$, $\theta_{\text{T}}$ and $\theta_{\text{R}}$. We define the channel parameter vector $\vect{\phi} = [\tau, \;\theta_{\text{T}}, \;\theta_{\text{R}}, \;\Re\{h\}, \;\Im\{h\}]\trans$, which contains the channel parameters and the real $\Re\{h\}$ and imaginary $\Im\{h\}$ parts of the channel gain as nuisance parameters. For observations under Gaussian noise, the CRLB states that the covariance matrix $\vect{C}_{\hat{\vect{\phi}}}$ of any unbiased estimator $\hat{\vect{\phi}}$ of $\vect{\phi}$ satisfies $\vect{C}_{\hat{\vect{\phi}}} - \vect{J}_{\vect{\phi}}^{-1}\succeq \vect{0}$, where $\succeq \vect{0}$ denotes positive semidefiniteness and $\vect{J}_{\vect{\phi}}\in\mathbb{R}^{5\times 5}$ is the Fisher information matrix (FIM), defined as
		\begin{IEEEeqnarray}{rCl}
			\vect{J}_{\vect{\phi}} &=& \frac{2}{\sigma_{\eta}^2} \sum_{p\in \mathcal{P}} \Re\left\{\frac{\partial \vect{m}\herm[p]}{\partial \vect{\phi}}\frac{\partial \vect{m}[p]}{\partial \vect{\phi}\trans}\right\}\label{eq:channel parameter FIM},
		\end{IEEEeqnarray}
		with $(\cdot)\herm$ denoting the conjugate transpose. The FIM $\vect{J}_{\tilde{\vect{\phi}}}\in\mathbb{R}^{5\times 5}$ and the CRLB for the position parameter vector $\tilde{\vect{\phi}} = [\vect{p}_{\text{R}},\;\alpha_{\text{R}},\; \Re\{h\}, \;\Im\{h\}]\trans$ can then be obtained as 
		\begin{IEEEeqnarray}{rCl}
			\vect{J}_{\tilde{\vect{\phi}}} &=& \vect{T}\vect{J}_{\vect{\phi}}\vect{T}\trans\label{eq:position parameter FIM},
		\end{IEEEeqnarray}
		where $\vect{T} = \partial \vect{\phi}\trans/\partial \tilde{\vect{\phi}} \in \mathbb{R}^{5\times 5}$.

	\section{Multi-Beam Signal Design}
		We would like to design the reference signal $\vect{x}[p],\;p\in\mathcal{P}$, subject to (s.t.) the Tx power constraint $\sum_{p\in\mathcal{P}}\left\|\vect{x}[p]\right\|_2^2 \leq P_{\text{T}}$, so as to improve the positioning accuracy of the Rx. We note that we could equivalently use energy as the limited resource. 
		
		The Tx excites $M_{\text{T}}\leq\left|\mathcal{P}\right|$ unit-norm beamforming directions $\vect{f}_k,\;k=1,\ldots,M_{\text{T}}$, 
		where $\left|\mathcal{P}\right|$ is the cardinality of $\mathcal{P}$. We denote with $\mathcal{P}_k$ the set of subcarriers allocated to $\vect{f}_k$. The reference signal can then be expressed as
		\begin{IEEEeqnarray}{rCl}
			\vect{x}[p] &=& \lambda_k[p]\vect{f}_k,\; p\in\mathcal{P}_k,
		\end{IEEEeqnarray}
		where 
		\begin{IEEEeqnarray}{rCl}
			\lambda_k[p] &=& P_{\text{T}}\sigma_k\sqrt{\gamma_{k,p}}\e^{\jj\mu_{k,p}}\label{eq:lambda_k[p]}
		\end{IEEEeqnarray} 
		is the symbol assigned to $\vect{f}_k$ at the $p$-th subcarrier, $\sigma_k^2$ is the fraction of $P_{\text{T}}$ allocated to $\vect{f}_k$, with $\sum_{k=1}^{M_{\text{T}}} \sigma_k^2 = 1$, $\gamma_{k,p}$ is the fraction of $\sigma_k^2$ allocated to the $p$-th subcarrier, $p\in\mathcal{P}_k$, with $\sum_{p\in\mathcal{P}_k}\gamma_{k,p} = 1$, and $\mu_{k,p}$ is the phase of $\lambda_k[p]$. In the following, we assume $\mathcal{P}_k$ and $\gamma_{k,p}$ to be given (see Remark~\ref{rem:rem1}).
		
		The position estimation accuracy is assessed in terms of the square position error bound (SPEB), defined as 
		\begin{IEEEeqnarray}{rCl}
			\text{SPEB}(\vect{q}, d, \theta_{\text{T}}) = \sum\nolimits_{i=1}^2 \vect{e}_i\trans\vect{J}_{\tilde{\vect{\phi}}}^{-1}(\vect{q}, d, \theta_{\text{T}})\vect{e}_i, \label{eq:SPEB definition}
		\end{IEEEeqnarray}
		where $\vect{q} = [\sigma_1^2,\ldots,\sigma_{M_{\text{T}}}^2]\trans\in\mathbb{R}^{M_{\text{T}}}$ and $\vect{e}_i$ is the vector of the appropriate size, whose $i$-th element is equal to $1$ and the rest of its elements are zero.
		
		We first study the case where $d$ and $\theta_{\text{T}}$ are perfectly known to the Tx. Building on this, we investigate the case where there is imperfect knowledge in the form of a probability density function (pdf). 
		
		\subsection{Perfect Knowledge}
			We distinguish between the case where the Tx is allowed to optimize both $\vect{f}_k$ and $\lambda_{k}[p],\; \;p\in\mathcal{P}_k,\;k=1,\ldots, M_{\text{T}}$, and the case where it uses a fixed set of beams, which we refer to as the beam codebook, and optimizes only $\lambda_k[p]$.
			
			\subsubsection{Optimal Beamforming Directions}
				In this case the optimization problem is
				\begin{IEEEeqnarray}{rCl}
					\min_{\vect{x}[p], p\in\mathcal{P}} \text{SPEB}(\vect{q}, d, \theta_{\text{T}})\quad\text{s.t. }  \sum\nolimits_{p\in\mathcal{P}}\left\|\vect{x}[p]\right\|_2^2 \leq P_{\text{T}}.
				\end{IEEEeqnarray}
				We define a couple of quantities that are useful for the presentation of the result: The effective bandwidth $\beta_k$ of the signal transmitted through the $k$-th beamforming direction is defined as \cite{KCS+18B}
				\begin{IEEEeqnarray}{rCl}
					\beta_k &=& \sqrt{\sum\nolimits_{p\in\mathcal{P}_k}\gamma_{k,p} \omega_p^2 - \Big(\sum\nolimits_{p\in\mathcal{P}_k}\gamma_{k,p} \omega_p\Big)^2 }
					\label{eq:effective baseband bandwidth}
					\IEEEeqnarraynumspace
				\end{IEEEeqnarray}
				and the array aperture function $\Xi_{\text{T}}\left(\theta_{\text{T}}\right)$ of the Tx array is defined as \cite{HSZ+16,KCS+18B}
				\begin{IEEEeqnarray}{rCl}
					\Xi_{\text{T}}\left(\theta_{\text{T}}\right) &=& \sqrt{\frac{1}{N_{\text{T}}} \sum\nolimits_{j=1}^{N_{\text{T}}}\left(d_{\text{T},j}\vect{u}_{\perp}\trans\left(\psi_{\text{T},j}\right)\vect{u}\left(\theta_{\text{T}}\right)\right)^2}.
					\IEEEeqnarraynumspace
				\end{IEEEeqnarray}
				
				\begin{theorem}
					\label{thm:optimal codebook}
					The optimal beamforming directions for SPEB minimization are the normalized array steering vector and its normalized derivative w.r.t. to $\theta_{\text{T}}$:
					\begin{IEEEeqnarray}{rCl}
						\vect{f}_{\text{opt},1}(\theta_{\text{T}}) &=& \frac{1}{\sqrt{N_{\text{T}}}}\vect{a}_{\text{T}}^*\left(\theta_{\text{T}}\right)\label{eq:f_1},\\
						\vect{f}_{\text{opt},2}(\theta_{\text{T}}) &=& \frac{c}{\omega_c\Xi_{\text{T}}\left(\theta_{\text{T}}\right)\sqrt{N_{\text{T}}}} \vect{D}_{\text{T}}^*\left(\theta_{\text{T}}\right) \vect{a}_{\text{T}}^*\left(\theta_{\text{T}}\right)\label{eq:f_2},
					\end{IEEEeqnarray}
					where $\vect{D}_{\text{T}} \left(\theta_{\text{T}}\right)$ is a diagonal matrix with $\left[\vect{D}_{\text{T}} \left(\theta_{\text{T}}\right)\right]_{j,j} = -\jj \frac{\omega_c}{c} d_{\text{T},j} \vect{u}_{\perp}\trans\left(\theta_{\text{T}}\right) \vect{u}(\psi_{\text{T},j}), j=1,\ldots,N_{\text{T}}$, with $\vect{u}_{\perp}(\psi) = \vect{u}\left(\psi -\pi/2\right)$ and $(\cdot)^*$ denoting the conjugate. The optimal power allocation is
					\begin{IEEEeqnarray}{rCl}
						\sigma_1^2(d, \theta_{\text{T}}) \hspace*{-0.021cm}=\hspace*{-0.023cm} \frac{\omega_c \Xi_{\text{T}}\left(\theta_{\text{T}}\right)}{\beta_1 d\hspace*{-0.023cm} + \hspace*{-0.023cm}\omega_c \Xi_{\text{T}}\left(\theta_{\text{T}}\right)}\label{eq:sigma_1^2 optimal},\; \sigma_2^2(d, \theta_{\text{T}})\hspace*{-0.023cm} =\hspace*{-0.023cm} 1\hspace*{-0.023cm} - \hspace*{-0.023cm} \sigma_1^2(d,\theta_{\text{T}}),\hspace*{-0.007cm}\label{eq:sigma_2^2 optimal}
						\IEEEeqnarraynumspace						
					\end{IEEEeqnarray}
					and the attained minimum is
					\begin{IEEEeqnarray}{rCl}
						\text{SPEB}_{\min}(d, \theta_{\text{T}}) &=& \frac{1}{g}\left(\frac{c}{\beta_1} + \frac{cd}{\omega_c \Xi_{\text{T}}\left(\theta_{\text{T}}\right)}\right)^2\label{eq:SPEB_min},
					\end{IEEEeqnarray}
					where $g = N_{\text{R}}N_{\text{T}}P_{\text{T}}\left|h\right|^2/\sigma_{\eta}^2$ is the Rx SNR.
				\end{theorem}
				\begin{IEEEproof}
					See Appendix \ref{sec:app thm1}.
				\end{IEEEproof}
				\begin{remark}
					\label{rem:rem1}
					From \eqref{eq:SPEB_min} we can observe that $\text{SPEB}_{\min}$ is a monotonically decreasing function of $\beta_1$, which depends on the fractions $\gamma_{1,p}$ of $\sigma_1^2$ allocated to the subcarriers in $\mathcal{P}_1$, that determine the waveform of the signal transmitted through $\vect{f}_{\text{opt},1}$. It can be shown that it is optimal to use only the edge subcarriers, i.e. $\mathcal{P}_1 = \{\min_{p\in\mathcal{P}}p,\;\max_{p\in\mathcal{P}}p\}$, with the power equally shared. However, as discussed in \cite{DJR+16}, this waveform choice results in higher sidelobes of the autocorrelation function of the signal transmitted through $\vect{f}_{\text{opt},1}$, which can degrade the delay estimation accuracy at low SNR. Since this topic is outside of the scope of the current work, we assume $\mathcal{P}_k$ and $\gamma_{k,p}$ to be given.
				\end{remark}
				
			\subsubsection{Beam Codebook}
				The Tx has to use a predefined codebook and optimize the power allocation $\vect{q}$ among the beams, which can lead to the selection of just a subset of them. Using \eqref{eq:channel parameter FIM} and \eqref{eq:position parameter FIM}, we find that $\vect{J}_{\tilde{\vect{\phi}}}(\vect{q}, d, \theta_{\text{T}})$ depends linearly on $\vect{q}$, i.e. $\vect{J}_{\tilde{\vect{\phi}}}(\vect{q}, d, \theta_{\text{T}})= \sum_{k=1}^{M_{\text{T}}}\sigma_k^2 \vect{J}_{\tilde{\vect{\phi}}}(\vect{e}_k, d, \theta_{\text{T}})$. The optimization problem can be expressed as 
				\begin{IEEEeqnarray}{rCl}
					\text{(G0)}&:&\; \min_{\vect{q}} \text{SPEB}(\vect{q}, d, \theta_{\text{T}})\quad \text{s.t.}\; \vect{q}\geq \vect{0},\; \vect{1}\trans\vect{q}\leq 1.
				\end{IEEEeqnarray}
				
				\begin{proposition}
					\label{prop:SDP}
					(G0) is equivalent to the following semidefinite program (SDP):
					\begin{IEEEeqnarray}{rCll}
						\text{(G1)}&:&\; \min_{\vect{q},\vect{B}} \trace(\vect{B})\;\; \text{s.t. }\hspace*{-0.05cm}& \begin{bmatrix}
							\vect{B} & \vect{E}\trans\\
							\vect{E} & \vect{J}_{\tilde{\vect{\phi}}}(\vect{q}, d, \theta_{\text{T}})
						\end{bmatrix} \succeq \vect{0},\nonumber\\
						&&&\vect{q}\geq \vect{0},\; \vect{1}\trans\vect{q}\leq 1
					\end{IEEEeqnarray}
					where $\vect{B}\in\mathbb{R}^{2\times 2},\;\vect{E} = \left[\vect{e}_1,\;\vect{e}_2\right]\in\mathbb{R}^{5\times 2}$ and $\trace(\cdot)$ is the trace operator.
				\end{proposition}
				\begin{IEEEproof}
					See \cite{LSZ+13}. 
				\end{IEEEproof}
				The formulation (G1) is important as it allows us to find the optimal power allocation in a computationally efficient way.
				
		\subsection{Imperfect Knowledge}
			The Tx has imperfect knowledge on the distance $d$ and the AOD $\theta_{\text{T}}$, which determine the Rx position, in the form of a joint pdf $p_{D,\Theta_{\text{T}}}\left(d,\theta_{\text{T}}\right)$. 
			In this section we stick to the beam codebook case and optimize the power allocation among the beams in terms of a metric for robust position estimation under the given imperfect knowledge on the channel parameters. An advantage of this approach for practical systems is that the Tx has to dynamically inform the Rx only about $M_T$ power allocation values, instead of 
			$M_{\text{T}}$ beamforming vectors of dimension $N_{\text{T}}$.
			In the following, we use two robustness-related metrics, namely the expected SPEB and the maximum SPEB.
			
			It can be shown that the SPEB can be expressed as
			\begin{IEEEeqnarray}{rCl}
				\hspace*{0.02cm}\text{SPEB}\left(\vect{q}, d, \theta_{\text{T}}\right) &=& \frac{1}{\left|h_0\right|^2 d_0^n}\left(\frac{d^n}{I_1\left(\vect{q},\theta_{\text{T}}\right)} + \frac{d^{n + 2}}{I_2\left(\vect{q},\theta_{\text{T}}\right)}\right)\label{eq:SPEB expression},
				\IEEEeqnarraynumspace
			\end{IEEEeqnarray}
			where $I_1$ and $I_2$ are some functions of $\vect{q}$ and $\theta_{\text{T}}$.
			From \eqref{eq:SPEB expression} we observe that the SPEB is a monotonically increasing function of $d$ and that, although that the SPEB depends on $h_0$ and $d_0$, the optimal power allocation for any of the considered metrics will be independent of them.
			
			\subsubsection{Expected SPEB Minimization}
				It follows from \eqref{eq:SPEB expression} that the expected SPEB can be expressed as
				\begin{IEEEeqnarray}{rCl}
					\mathbb{E}_{D,\Theta_{\text{T}}}&&\left[\text{SPEB}\left(\vect{q}, D, \Theta_{\text{T}}\right)\right] \nonumber\\
					=&&\; \E{w\left(\Theta_{\text{T}}\right)\text{SPEB}\left(\vect{q}, \bar{d}\left(\Theta_{\text{T}}\right), \Theta_{\text{T}}\right)}{\Theta_{\text{T}}},
				\end{IEEEeqnarray}
				where 
				\begin{IEEEeqnarray}{rCl}
					\bar{d}\left(\theta_{\text{T}}\right) &=& \sqrt{\E{d^{n+2}}{D|\Theta_{\text{T}}=\theta_{\text{T}}}/\E{d^{n}}{D|\Theta_{\text{T}}=\theta_{\text{T}}}}\\
					w\left(\theta_{\text{T}}\right) &=& \E{d^{n}}{D|\Theta_{\text{T}}=\theta_{\text{T}}}/\bar{d}^n\left(\theta_{\text{T}}\right),
				\end{IEEEeqnarray}
				with $\E{f(X)}{X}$ denoting the expected value of the function $f(X)$ over the distribution of $X$.
				
				In order to solve the optimization problem in hand, we choose a discretization $(\theta_{\text{T},l}, P_{\Theta_{\text{T}}}\left(\theta_{\text{T},l}\right)), l = 1,\ldots, N_{\theta}$, of the marginal distribution $p_{\Theta_{\text{T}}}$, where $N_{\theta}$ is the number of discretization points and $P_{\Theta_{\text{T}}}$ is the resulting probability mass function (pmf). Then, our optimization problem becomes
				\begin{IEEEeqnarray}{rCl}
					\min_{\vect{q}}&& \sum_{l=1}^{N_{\theta}} P_{\Theta_{\text{T}}}\left(\theta_{\text{T},l}\right) w\left(\theta_{\text{T},l}\right)\text{SPEB}\left(\vect{q}, \bar{d}\left(\theta_{\text{T},l}\right), \theta_{\text{T},l}\right)\nonumber\\
					&& \text{s.t. } \vect{q}\geq \vect{0},\; \vect{1}\trans\vect{q}\leq 1\label{eq:expectation minimization optimization problem standard form}
					\IEEEeqnarraynumspace
				\end{IEEEeqnarray}
				Similar to (G0) and (G1), the problem in \eqref{eq:expectation minimization optimization problem standard form} is equivalent to the following SDP
				\begin{IEEEeqnarray}{rCl}
					&&\min_{\vect{q},\vect{B}_1,\ldots, \vect{B}_{N_{\theta}}} \sum_{l=1}^{N_{\theta}} P_{\Theta_{\text{T}}}\left(\theta_{\text{T},l}\right) w\left(\theta_{\text{T},l}\right) \trace\left(\vect{B}_l\right)\nonumber\\
					&&\;\;\text{s.t. } \begin{bmatrix}
					\vect{B}_l & \vect{E}\trans\\
					\vect{E} & \vect{J}_{\tilde{\vect{\phi}}}\left(\vect{q}, \bar{d}\left(\theta_{\text{T},l}\right), \theta_{\text{T},l}\right)
					\end{bmatrix} \succeq \vect{0},\;l=1,\ldots,N_{\theta}\nonumber\\
					&&\quad\;\;\;\;\; \vect{q}\geq \vect{0},\; \vect{1}\trans\vect{q}\leq 1,\label{eq:minexp SDP}
				\end{IEEEeqnarray}
				which can be solved efficiently.
				
			\subsubsection{Maximum SPEB Minimization}
				Since the SPEB is a monotonically increasing function of $d$, it is straightforward that
				\begin{IEEEeqnarray}{rCl}
					\max_{\substack{\left(d, \theta_{\text{T}}\right)\in\\\supp p_{D,\Theta_{\text{T}}}}}\hspace*{-0.15cm}\text{SPEB}\left(\vect{q}, d, \theta_{\text{T}}\right) = \hspace*{-0.05cm}\max_{\substack{\theta_{\text{T}}\in\\\supp p_{\Theta_{\text{T}}}}} \hspace*{-0.15cm}\text{SPEB}\left(\vect{q}, d_{\max}\left(\theta_{\text{T}}\right), \theta_{\text{T}}\right),\nonumber
				\end{IEEEeqnarray}
				where 	
				$d_{\max}\left(\theta_{\text{T}}\right) = \max_{d\in\supp p_{D|\theta_{\text{T}}}} d$. Following the same process as before, we can solve the power allocation problem efficiently using its equivalent SDP reformulation. 
				
	\section{Numerical Results}
		\subsection{System Parameters}
			We consider a system operating at $f_c = \SI{38}{\giga\hertz}$, with $N = 4096$ subcarriers, subcarrier spacing $f_{\text{s}}/N = \SI{30}{\kilo\hertz}$ and $\mathcal{P} = \{-1197,-1191,...,1191,1197\}$, with $|\mathcal{P}|=400$. The Tx and Rx are equipped with ULAs with $N_{\text{T}} =  32$ and $N_{\text{R}} = 4$. The noise variance is $\sigma_{\eta}^2 = 2 \cdot 10^{0.1 N_0} f_{\text{s}}$, where $N_0=\SI{-170}{\dBm\per\hertz}$ is the noise power spectral density per dimension. The clock synchronization error standard deviation is $\sigma_{\text{clk}} = 0.25/f_{\text{s}}$, so that $c \sigma_{\text{clk}} \approx \SI{0.61}{\meter}$. We also set $n=2$, $d_0 =\SI{1}{\meter}$, $h_0 = c/(4\pi f_c)$ and $P_{\text{T}}=\SI{0}{\decibel}$, which results in a transmission power of $\SI{4}{\dBm}$ per time-domain sample.
			
		\subsection{Considered Codebooks and Power Allocation Schemes}
			\label{sec:schemes description}
			In the following results we compare the following codebooks and power allocation schemes in terms of the position error bound $\text{PEB} = \sqrt{\text{SPEB}}$:
			\begin{itemize}
				\item Optimal beamforming directions \eqref{eq:f_1}-\eqref{eq:f_2} $\vect{f}_{\text{opt}, 1}(\mu_{\theta_{\text{T}}})$ and $\vect{f}_{\text{opt}, 2}(\mu_{\theta_{\text{T}}})$, with power allocation \eqref{eq:sigma_1^2 optimal} $\sigma_1^2(\mu_d, \mu_{\theta_{\text{T}}})$ ("opt-perf"), where $\mu_{\theta_{\text{T}}} = \E{\Theta_{\text{T}}}{\theta_{\text{T}}}$ and $\mu_d = \E{D}{D}$.
				\item Discrete Fourier Transform (DFT) codebook, 
				with power allocation minimizing the expected ("DFT minexp") or the maximum ("DFT minmax") PEB.
				We also consider a heuristic scheme ("DFT uni"), where the available power is uniformly allocated to the subset $\mathcal{B}_{\text{uni}}$ of the DFT beams maximizing the projection to $\vect{f}_{\text{opt}, 1}$:
				\begin{IEEEeqnarray}{rCl}
					\mathcal{B}_{\text{uni}} &=& \cup_{l=1}^{N_{\theta}} \Big\{\argmax_{k=1,\ldots,N_{\text{T}}}|\vect{f}_{\text{opt}, 1}\trans(\theta_{\text{T},l})\vect{f}_k|\Big\}. 
				\end{IEEEeqnarray}  
				\item  Motivated by Theorem~\ref{thm:optimal codebook}, observing that for a ULA the $k$-th beam of the DFT codebook is collinear with $\vect{a}_{\text{T}}^*\left(\theta_k\right)$, where $\sin(\theta_k) = 2(k-1)/N_{\text{T}} - 1$, we augment the DFT codebook with the unit-norm beams that are collinear with $\vect{D}_{\text{T}}^*\left(\theta_k\right)\vect{a}_{\text{T}}^*\left(\theta_k\right),\; k=1,\ldots, N_{\text{T}}$. We consider similar power allocation schemes as for the DFT codebook and refer to them as "DFT\&D minexp", "DFT\&D minmax" and "DFT\&D uni".
				\item Wost case-CRLB-minimizing codebook following~\cite{GWS18} ("AOD-opt") of size $N_T$. The optimization of the power allocation has negligible impact on its performance, hence we use uniform power allocation over its beams.
			\end{itemize}
			For all codebooks, we allocate subcarriers to the beams in an interleaved manner and uniformly allocate the power over the subcarriers within the resulting subsets, i.e. $\mathcal{P}_k =\{\min_{p\in\mathcal{P}}p + k, \min_{p\in\mathcal{P}}p + k + M_{\text{T}}, \ldots\},$ and $\gamma_{k,p} = 1/|\mathcal{P}_k|$.
		
		\subsection{Coarse Prior Knowledge of \texorpdfstring{$d$}{d} and \texorpdfstring{$\theta_{\text{T}}$}{thetaT}}
			\label{sec:coarse knowledge results}
			
			In this example, we assume that, through some preceding estimation process, the Tx has acquired coarse knowledge of the channel parameters in the form of a distribution over them. $\theta_{\text{T}}$ follows a von Mises (wrapped Gaussian) distribution $p_{\theta_{\text{T}}}(\theta_{\text{T}}) = \e^{\cos(\theta_{\text{T}} - \mu_{\theta_{\text{T}}})/\sigma_{\theta_{\text{T}}}^2}/(2\pi I_0(1/\sigma_{\theta_{\text{T}}}^2))$, where $I_0(\cdot)$ is the zero-order modified Bessel function of the first kind, $\mu_{\theta_{\text{T}}} = \ang{25}$ and $\sigma_{\theta_{\text{T}}} = \ang{7.5}$. 
			We truncate $p_{\theta_{\text{T}}}$ at $\mu_{\theta_{\text{T}}}\pm 2\sigma_{\theta_{\text{T}}}$, so that it has a bounded support, which is necessary in order to get reasonable results for the "minmax" approaches, and discretize it using a trapezoidal rule with $N_{\theta}=127$.
			We also assume that $d$ is independent of $\theta_{\text{T}}$ and follows a Gaussian distribution with mean $\mu_d = \SI{35}{\meter}$ and standard deviation $\sigma_d = \SI{7.5}{\meter}$, which we truncate at $\mu_d \pm 2\sigma_d$.
			
\begin{figure}
				\centering
				\begin{adjustbox}{scale=0.99}
					\includegraphics[]{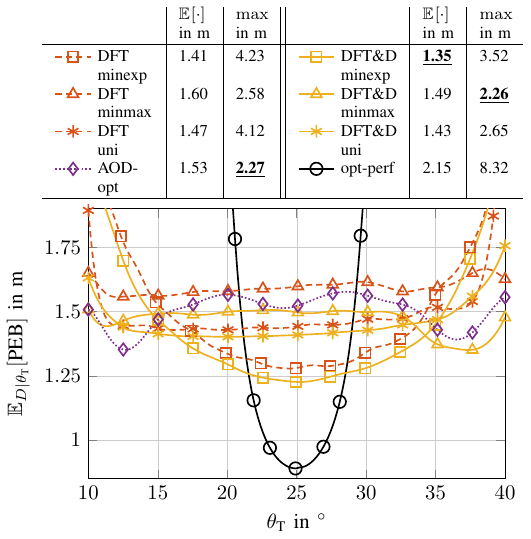}
				\end{adjustbox}
				\caption{$\E{\text{PEB}}{D|\theta_{\text{T}}}$ vs $\theta_{\text{T}}$ for the schemes described in Sec.~\ref{sec:schemes description}.}
				\label{fig:PEB_vs_theta_T}
			\end{figure}

\begin{figure}
				\centering
				\subfloat[opt-perf]{\begin{adjustbox}{scale=0.71}
						\includegraphics[]{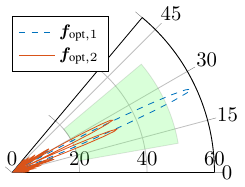}
				\end{adjustbox}}\hspace*{-0.35cm}
				\subfloat[DFT minexp]{\begin{adjustbox}{scale=0.71}
						\includegraphics{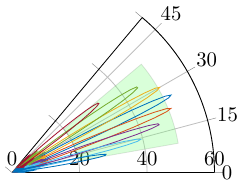}
				\end{adjustbox}}\hspace*{-0.35cm}
				\subfloat[DFT minmax]{\begin{adjustbox}{scale=0.71}
						\includegraphics{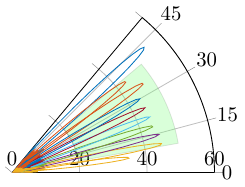}
				\end{adjustbox}}
				\caption{Optimal beamforming directions and power-optmized DFT codebook. The shaded region represents the support of $p_{D\Theta_{\text{T}}}$.}
				\label{fig:VC_polar}
			\end{figure}

			In Fig.~\ref{fig:PEB_vs_theta_T} we plot the average PEB over the distance $\E{\text{PEB}}{D|\theta_{\text{T}}}$ as a function of $\theta_{\text{T}}$ for all codebooks and power allocation schemes listed in Sec.~\ref{sec:schemes description}. 
			The resulting beam patterns for three of the considered configurations are shown in Fig.~\ref{fig:VC_polar}. In Fig.~\ref{fig:VC_polar}(a) we see that $\vect{f}_{\text{opt}, 2}(\mu_{\theta_{\text{T}}})$ forms two equally strong lobes surrounding the beam formed by $\vect{f}_{\text{opt}, 1}(\mu_{\theta_{\text{T}}})$. 

			As seen in Fig.~\ref{fig:VC_polar}(b), for expectation minimization with the DFT codebook, most of the available power is allocated to the beams near $\mu_{\theta_{\text{T}}}$. For maximum minimization (see Fig.~\ref{fig:VC_polar}(c)), the power allocation is more asymmetric, especially at the edges of the angular support. This is due to the fact that $\mu_{\theta_{\text{T}}}+2\sigma_{\theta_{\text{T}}}$ is closer to the array's endfire, and, therefore, harder to estimate, whereas $\mu_{\theta_{\text{T}}}-2\sigma_{\theta_{\text{T}}}$ is close to its broadside.
		
			In Fig.~\ref{fig:PEB_vs_theta_T} we can see that the sharp optimal beamforming vectors directed to $\mu_{\theta_{\text{T}}}$ result in the minimum error, but their performance is sensitive to deviations of $\theta_{\text{T}}$ from $\mu_{\theta_{\text{T}}}$. As expected, the performance of "minmax" schemes is more flat over the angle, but "minexp" schemes can provide a lower PEB for a wide range of angles.
			The DFT\&D codebook with the appropriate power allocation scheme has the best performance for both considered metrics. Comparisons with an oversampled DFT codebook, not included here for brevity, have shown that its superiority cannot only be attributed to its denser spatial sampling, but also to the shape of its beams. Finally, we note that the AOD-opt
			codebook, although not accounting for the distance component of the target's position, exhibits very good performance, especially in terms of maximum minimization.
			
		\subsection{Downlink Vehicle-to-everything (V2X) Scenario}
\begin{figure}
				\centering
				\subfloat[Geometry]{\begin{adjustbox}{scale=0.88}
						\includegraphics{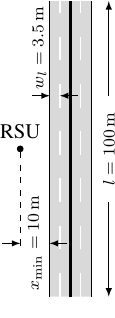}\hspace*{0.5cm}
				\end{adjustbox}}\quad
				\subfloat[DFT minexp]{\begin{adjustbox}{scale=0.63}
						\includegraphics{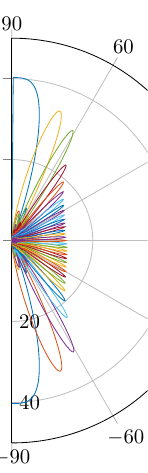}\hspace*{0.5cm}
				\end{adjustbox}}\quad
				\subfloat[DFT minmax]{\begin{adjustbox}{scale=0.63}
						\includegraphics{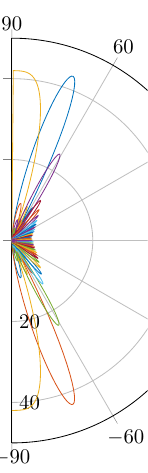}\hspace*{0.5cm}
				\end{adjustbox}}
				\caption{V2X scenario geometry and power-optmized DFT codebook.}
				\label{fig:VD_polar}
			\end{figure}

			In this setup, which is shown in Fig.~\ref{fig:VD_polar}(a), a road side unit (RSU) serves vehicles in a $l=\SI{100}{\meter}$ long road segment with 4 lanes, which have a width of $w_l = \SI{3.5}{\meter}$. 
			The RSU is located midway of the segment, $x_{\min}=\SI{10}{m}$ away from its left edge. The vehicles are uniformly distributed on the road, resulting in a $p_{D,\Theta_{\text{T}}}$ with bounded 
			support. Again, we discretize $p_{\Theta_{\text{T}}}$ using a trapezoidal rule with $N_{\theta}=127$.
			We consider all configurations described in Sec.~\ref{sec:schemes description}, apart from "opt-perf", which is meaningless for the wide range of angles involved. The resulting optimal power allocation for the DFT codebook is depicted in Figs.~\ref{fig:VD_polar}(b) and (c), where we can see that for both optimization strategies, and especially for minmax, the edge beams get most of the available power.
			
\begin{figure}
				\centering
				\begin{adjustbox}{scale=0.91}
				\includegraphics{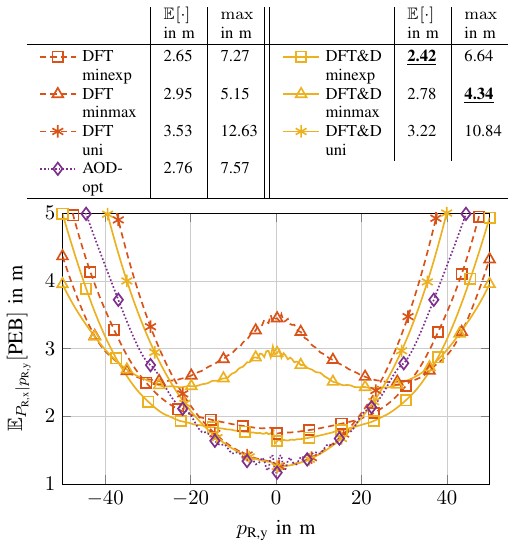}
				\end{adjustbox}
				\caption{$\E{\text{PEB}}{P_{\text{R,x}}|p_{\text{R,y}}}$ vs $p_{\text{R,y}}$ for the schemes described in Sec.~\ref*{sec:schemes description}.}
				\label{fig:PEB_vs_x}
			\end{figure}

			In Fig.~\ref{fig:PEB_vs_x} we plot the average error over the road's width $\E{\text{PEB}}{P_{\text{R,x}}|p_{\text{R,y}}}$ as a function of 
			$p_{\text{R,y}}$.
			Uniform power allocation can offer a low PEB near the middle of the road segment, but it results in the worst performance among all configurations in terms of the considered metrics. The DFT\&D codebook can offer both the lowest $\E{\text{PEB}}{}$ and $\min\text{PEB}$ with the appropriate power allocation strategy. In this scenario the AOD-opt codebook does not perform as well as in the example of Sec.~\ref{sec:coarse knowledge results},
			particularly in terms of minimization of the maximum PEB
			for the following reason: here, the support of $p_{D|\theta_{\text{T}}}$, and therefore the maximum distance, is different for each $\theta_{\text{T}}$. 
			Let $\theta_{\text{T},1}$ and $\theta_{\text{T},2}$ be two angles with the same uncertainty, as expressed by the square roots of their respective AOD CRLBs.
			If $d_{\max}(\theta_{\text{T},1})>d_{\max}(\theta_{\text{T},2})$, it will be preferable to spend more energy in the direction of $\theta_{\text{T},1}$, as, for a given angular uncertainty, the position uncertainty grows linearly with $d$~\cite{AZA+18}. This is not accounted for by a codebook optimized for AOD estimation.
			On top of that, by employing a distance-dependent path loss model, the Tx can take into account the different worst-case SNR for different angles.
			
	\section{Conclusion}
		In this work, we discussed how prior knowledge of the channel parameters can enhance the  performance of single-anchor positioning in a LOS scenario. With perfect knowledge of the distance and AOD, we derived the optimal beamforming directions and power allocation. With imperfect prior knowledge, we considered a beam codebook and optimized the power allocation among its beams w.r.t. two different robustness metrics. We also identified a simple augmented DFT codebook, which, combined with appropriate power allocation optimization, offers equal or even better positioning accuracy than existing approaches that incur higher communication 
		and computational overhead. 
			
	\appendix[Proof of Theorem~\ref{thm:optimal codebook}]

		\label{sec:app thm1}
		With the Tx array's centroid $\bar{\vect{p}}_{\text{T}} = \sum_{j=1}^{N_T} \vect{p}_{\text{T},j}$ chosen as its reference point we have $
			\vect{a}_{\text{T}}\herm\left(\theta_{\text{T}}\right) \vect{D}_{\text{T}}\left(\theta_{\text{T}}\right) \vect{a}_{\text{T}}\left(\theta_{\text{T}}\right) = 0$,
		that is $\vect{f}_{\text{opt},1}(\theta_{\text{T}})$ and $\vect{f}_{\text{opt},2}(\theta_{\text{T}})$, as defined in \eqref{eq:f_1}-\eqref{eq:f_2}, are orthogonal. Hence, we write
		\begin{IEEEeqnarray}{rCl}
			\vect{x}[p] &=& \left[\vect{v}_1(\theta_{\text{T}}),\; \vect{v}_2(\theta_{\text{T}}),\;\vect{W}(\theta_{\text{T}})\right]\vect{\zeta}[p]\label{eq:reference signal as linear combination of basis},
		\end{IEEEeqnarray}
		where $\vect{v}_1(\theta_{\text{T}}) = \vect{f}_{\text{opt},1}(\theta_{\text{T}}),\; \vect{v}_2(\theta_{\text{T}}) = \vect{f}_{\text{opt},2}(\theta_{\text{T}})$ and $\vect{W}(\theta_{\text{T}})\in\mathbb{C}^{N_{\text{T}}\times N_{\text{T}} - 2}$ is a set of vectors which span the subspace of $\mathbb{C}^{N_{\text{T}}}$ that is orthogonal to $\vect{v}_1(\theta_{\text{T}})$ and $\vect{v}_2(\theta_{\text{T}})$. This imposes no restrictions on the reference signal, as we have just expressed it as a linear combination of basis vectors of $\mathbb{C}^{N_{\text{T}}}$. Computing $\vect{J}_{\vect{\phi}}$ according to \eqref{eq:channel parameter FIM}, we find that transmission on the subspace spanned by $\vect{W}(\theta_{\text{T}})$ does not contribute to the Fisher information. Therefore, all the energy should be allocated to the subspace spanned by $\vect{v}_1(\theta_{\text{T}})$ and $\vect{v}_2(\theta_{\text{T}})$.
		
		After some computations we get
		\begin{IEEEeqnarray}{rCl}
			\text{SPEB} &=& \frac{c^2}{J_{\tau}' - \big(J_{\tau\theta_{\text{T}}}'\big)^2/J_{\theta_{\text{T}}}'} + \frac{ d^2}{J_{\theta_{\text{T}}}' - \big(J_{\tau\theta_{\text{T}}}'\big)^2/J_{\tau}'} \label{eq:app_SPEB},
		\end{IEEEeqnarray}
		where
		\begin{IEEEeqnarray}{rCl}
			J_{\tau}' &=& \frac{g}{P_{\text{T}}}  \left(\sum\nolimits_{p\in\mathcal{P}}\left|\zeta_1[p]\right|^2 \omega_p^2\hspace*{-0.01cm} - \hspace*{-0.01cm}\left(\sum\nolimits_{p\in\mathcal{P}}\left|\zeta_1[p]\right|^2 \omega_p\right)^2 \right)\label{eq:J_tau'},\nonumber\\
			J_{\theta_{\text{T}}}' &=& \frac{g}{P_{\text{T}}} \frac{\omega_{\text{c}}^2}{c^2}\Xi_{\text{T}}^2\left(\theta_{\text{T}}\right) \hspace*{-0.023cm} \left(\hspace*{-0.019cm}\sum_{p\in\mathcal{P}}\left|\zeta_2[p]\right|^2\hspace*{-0.019cm} -\hspace*{-0.019cm} \frac{\left|\sum_{p\in \mathcal{P}}\zeta_1[p]\zeta_2^*[p]\right|^2}{\sum_{p\in\mathcal{P}}\left|\zeta_1[p]\right|^2} \hspace*{-0.019cm}\right)\hspace*{-0.01cm}\label{eq:J_theta'},\nonumber\\
			J_{\tau\theta_{\text{T}}}' &=&  \frac{g}{P_{\text{T}}} \frac{\omega_{\text{c}}}{c} \Xi_{\text{T}}\left(\theta_{\text{T}}\right) \sum\nolimits_{p\in\mathcal{P}} \Im\left\{\zeta_1[p]\zeta_2^*[p]\right\} \tilde{\omega}_p,\label{eq:J_tautheta'}
		\end{IEEEeqnarray}
		with $\Im{\cdot}$ denoting the imaginary part, $\zeta_i[p]$ being the $i$-th element of $\vect{\zeta}[p]$ and
		\begin{IEEEeqnarray}{rCl}
			\tilde{\omega}_p &=& \frac{\sum_{p'\in\mathcal{P}}\left|\zeta_1[p']\right|^2\left(\omega_p -\omega_{p'}\right)}{\sum_{p'\in\mathcal{P}}\left|\zeta_1[p']\right|^2}\nonumber.
		\end{IEEEeqnarray}
		From \eqref{eq:app_SPEB}-\eqref{eq:J_tautheta'} we conclude that in order to minimize the $\text{SPEB}$ we have to choose the sequences $\zeta_1[p], \zeta_2[p]$ such that
		\begin{IEEEeqnarray}{rCl}
			\sum\nolimits_{p\in \mathcal{P}}\zeta_1[p]\zeta_2^*[p] &=& 0,\;\;\sum\nolimits_{p\in \mathcal{P}}\Im\left\{\zeta_1[p]\zeta_2^*[p]\right\} \tilde{\omega}_p = 0.
			\IEEEeqnarraynumspace
		\end{IEEEeqnarray}
		A straightforward way to achieve that is to choose
		\begin{IEEEeqnarray}{rCl}
			\zeta_k[p] &=& \begin{cases}
				\lambda_k[p],\;& p\in\mathcal{P}_k,\\
				0,\; & \text{otherwise},
			\end{cases}\label{eq:zeta_k[p] optimal}
		\end{IEEEeqnarray}
		where $\mathcal{P}_k,\;k=1,2$, are disjoint subsets of $\mathcal{P}$ and $\lambda_{k}[p]$ defined in \eqref{eq:lambda_k[p]}, so that $\zeta_1[p]\zeta_2[p]=0,\;\forall p\in\mathcal{P}$. We have hence shown that $\vect{f}_{\text{opt},1}(\theta_{\text{T}})$ and $\vect{f}_{\text{opt},2}(\theta_{\text{T}})$ are an optimal codebook for SPEB minimization. Using \eqref{eq:zeta_k[p] optimal} and \eqref{eq:lambda_k[p]}, we write \eqref{eq:app_SPEB} as
		\begin{IEEEeqnarray}{rCl}
			\text{SPEB}(\sigma_1^2, d, \theta_{\text{T}}) &=& \frac{c^2}{g \sigma_1^2\beta_1^2} + \frac{ c^2 d^2}{g \sigma_2^2\omega_{\text{c}}^2\Xi_{\text{T}}^2\left(\theta_{\text{T}}\right)}\label{eq:SPEB with optimal codebook}.
		\end{IEEEeqnarray}
		Replacing $\sigma_2^2 = 1 - \sigma_1^2$ in~\eqref{eq:SPEB with optimal codebook}, as dictated by the power constraint, we find that the SPEB is a convex function of $\sigma_1^2$. Setting its derivative w.r.t. $\sigma_1^2$ to zero gives us \eqref{eq:sigma_1^2 optimal}.
		
	\bibliographystyle{IEEEtran}
	\bibliography{IEEEabrv,5G_Downlink_Multi-Beam_Signal_Design_for_LOS_positioning}

\begin{thebibliography}{10}
\providecommand{\url}[1]{#1}
\csname url@samestyle\endcsname
\providecommand{\newblock}{\relax}
\providecommand{\bibinfo}[2]{#2}
\providecommand{\BIBentrySTDinterwordspacing}{\spaceskip=0pt\relax}
\providecommand{\BIBentryALTinterwordstretchfactor}{4}
\providecommand{\BIBentryALTinterwordspacing}{\spaceskip=\fontdimen2\font plus
\BIBentryALTinterwordstretchfactor\fontdimen3\font minus
  \fontdimen4\font\relax}
\providecommand{\BIBforeignlanguage}[2]{{%
\expandafter\ifx\csname l@#1\endcsname\relax
\typeout{** WARNING: IEEEtran.bst: No hyphenation pattern has been}%
\typeout{** loaded for the language `#1'. Using the pattern for}%
\typeout{** the default language instead.}%
\else
\language=\csname l@#1\endcsname
\fi
#2}}
\providecommand{\BIBdecl}{\relax}
\BIBdecl

\bibitem{WHK+16}
K.~Witrisal, S.~Hinteregger, J.~Kulmer, E.~Leitinger, and P.~Meissner,
  ``High-accuracy positioning for indoor applications: {RFID}, {UWB}, 5{G}, and
  beyond,'' in \emph{IEEE 10th Int. Conf. RFID}, May 2016, pp. 1--7.

\bibitem{WSD+17}
H.~Wymeersch, G.~Seco-Granados, G.~Destino, D.~Dardari, and F.~Tufvesson,
  ``5{G} mm{W}ave positioning for vehicular networks,'' \emph{IEEE Wireless
  Commun.}, vol.~24, no.~6, pp. 80--86, Dec. 2017.

\bibitem{HSZ+16}
Y.~Han, Y.~Shen, X.~P. Zhang, M.~Z. Win, and H.~Meng, ``Performance limits and
  geometric properties of array localization,'' \emph{{IEEE} Trans. Inf.
  Theory}, vol.~62, no.~2, pp. 1054--1075, Feb. 2016.

\bibitem{MB19}
R.~{Mendrzik} and G.~{Bauch}, ``Position-constrained stochastic inference for
  cooperative indoor localization,'' \emph{{IEEE} Trans. Signal Inf. Process.
  Netw.}, pp. 1--1, Feb. 2019.

\bibitem{SAH+14}
A.~L. {Swindlehurst}, E.~{Ayanoglu}, P.~{Heydari}, and F.~{Capolino},
  ``Millimeter-wave massive {MIMO}: the next wireless revolution?''
  \emph{{IEEE} Commun. Mag.}, vol.~52, no.~9, pp. 56--62, Sep. 2014.

\bibitem{SGD+15}
A.~Shahmansoori, G.~E. Garcia, G.~Destino, G.~Seco-Granados, and H.~Wymeersch,
  ``5{G} position and orientation estimation through millimeter wave {MIMO},''
  in \emph{IEEE GLOBECOM Workshops (GC Wkshps)}, Dec. 2015, pp. 1--6.

\bibitem{AZA+18}
Z.~Abu{-}Shaban, X.~Zhou, T.~D. Abhayapala, G.~Seco{-}Granados, and
  H.~Wymeersch, ``Error bounds for uplink and downlink {3D} localization in
  {5G} {mmWave} systems,'' \emph{{IEEE} Trans. Wireless Commun.}, vol.~17,
  no.~8, pp. 4939--4954, Aug. 2018.

\bibitem{GSK+18}
G.~E. {Garcia}, G.~{Seco-Granados}, E.~{Karipidis}, and H.~{Wymeersch},
  ``Transmitter beam selection in millimeter-wave {MIMO} with in-band
  position-aiding,'' \emph{{IEEE} Trans. Wireless Commun.}, vol.~17, no.~9, pp.
  6082--6092, Sep. 2018.

\bibitem{FCW+18}
\BIBentryALTinterwordspacing
A.~Fascista, A.~Coluccia, H.~Wymeersch, and G.~Seco-Granados, ``Millimeter-wave
  downlink positioning with a single-antenna receiver,'' \emph{CoRR}, vol.
  abs/1811.11586, 2018. [Online]. Available:
  \url{http://arxiv.org/abs/1811.11586}
\BIBentrySTDinterwordspacing

\bibitem{GWS+16}
N.~{Garcia}, H.~{Wymeersch}, E.~G. {Ström}, and D.~{Slock}, ``Location-aided
  mm-wave channel estimation for vehicular communication,'' in \emph{IEEE 17th
  Int. Workshop on Signal Processing Advances in Wireless Communications
  (SPAWC)}, Jul. 2016, pp. 1--5.

\bibitem{KDD+17}
R.~{Koirala}, B.~{Denis}, D.~{Dardari}, and B.~{Uguen}, ``Localization bound
  based beamforming optimization for multicarrier mm{W}ave {MIMO},'' in
  \emph{14th Workshop on Positioning, Navigation and Communications (WPNC)},
  Oct. 2017, pp. 1--6.

\bibitem{GWS18}
N.~{Garcia}, H.~{Wymeersch}, and D.~T.~M. {Slock}, ``Optimal precoders for
  tracking the {AoD} and {AoA} of a mm{W}ave path,'' \emph{{IEEE} Trans. Signal
  Process.}, vol.~66, no.~21, pp. 5718--5729, Nov. 2018.

\bibitem{KDU+18}
R.~{Koirala}, B.~{Denis}, B.~{Uguen}, D.~{Dardari}, and H.~{Wymeersch},
  ``Localization optimal multi-user beamforming with multi-carrier mm{W}ave
  {MIMO},'' in \emph{IEEE Int. Symposium on Personal, Indoor and Mobile Radio
  Commun. (PIMRC)}, Sep. 2018, pp. 1--7.

\bibitem{ZZS18}
H.~{Zhao}, L.~{Zhang}, and Y.~{Shen}, ``On the optimal beamspace design for
  direct localization systems,'' in \emph{IEEE International Conference on
  Communications (ICC)}, May 2018, pp. 1--6.

\bibitem{KCS+18B}
\BIBentryALTinterwordspacing
A.~Kakkavas, M.~H. {Casta{\~{n}}eda Garc{\'{\i}}a}, R.~A. Stirling{-}Gallacher,
  and J.~A. Nossek, ``Performance limits of single-anchor mm-{W}ave
  positioning,'' \emph{CoRR}, vol. abs/1808.08116, 2018. [Online]. Available:
  \url{http://arxiv.org/abs/1808.08116}
\BIBentrySTDinterwordspacing

\bibitem{DJR+16}
A.~{Dammann}, T.~{Jost}, R.~{Raulefs}, M.~{Walter}, and S.~{Zhang},
  ``Optimizing waveforms for positioning in 5{G},'' in \emph{IEEE 17th Int.
  Workshop on Signal Processing Advances in Wireless Commun. (SPAWC)}, Jul.
  2016, pp. 1--5.

\bibitem{LSZ+13}
W.~W. {Li}, Y.~{Shen}, Y.~J. {Zhang}, and M.~Z. {Win}, ``Robust power
  allocation for energy-efficient location-aware networks,'' \emph{IEEE/ACM
  Trans. Networking}, vol.~21, no.~6, pp. 1918--1930, Dec. 2013.

\end{thebibliography}
	
\end{document}